%% file: Template.tex
\documentclass{article}
\usepackage{spconf,amsmath,graphicx, multirow}
\usepackage{multicol}
\usepackage{hyperref}
\usepackage[dvipsnames]{xcolor}

\usepackage{soul}

\title{Reducing Spelling Inconsistencies in Code-Switching ASR using Contextualized CTC Loss}
  
\name{\parbox{\linewidth}{\centering Burin Naowarat$^{\star}$ \qquad Thananchai Kongthaworn$^{\star}$ \qquad Korrawe Karunratanakul$^{\ddagger}$ \\ Sheng Hui Wu $^{\dagger}$ \qquad Ekapol Chuangsuwanich$^{\star}$}}

\address{
    $^{\star}$ Department of Computer Engineering, Chulalongkorn University, Thailand \\
    $^{\ddagger}$ ETH Zurich, Switzerland ~     $^{\dagger}$ NewEra AI Robotics, Taiwan \\\\
    \ninept{\{6270145221, 6370120621\}@student.chula.ac.th \qquad kkarunrat@inf.ethz.ch \qquad areoll@neweraai.com \qquad ekapolc@cp.eng.chula.ac.th}
    }
    
\begin{document}
\onecolumn
\noindent IEEE Copyright Notice:

© 2020 IEEE.  Personal use of this material is permitted. Permission from IEEE must be obtained for all other uses, in any current or future media, including reprinting/republishing this material for advertising or promotional purposes, creating new collective works, for resale or redistribution to servers or lists, or reuse of any copyrighted component of this work in other works.
\clearpage

\twocolumn
\ninept
\maketitle
\begin{abstract}
\input{tex_files/abstract}
\end{abstract}
\begin{keywords}
code-switching, end-to-end speech recognition, context prediction, connectionist temporal classification
\end{keywords}
\input{tex_files/intro}
\input{tex_files/background}
\input{tex_files/method}
\input{tex_files/dataset}
\input{tex_files/exp}
\input{tex_files/conclusion}

\vfill\pagebreak

\bibliographystyle{IEEEbib}
\bibliography{strings,refs}

\end{document}

%% file: tex_files/abstract.tex
Code-Switching (CS) remains a challenge for Automatic Speech Recognition (ASR), especially character-based models. With the combined choice of characters from multiple languages, the outcome from character-based models suffers from phoneme duplication, resulting in language-inconsistent spellings. We propose Contextualized Connectionist Temporal Classification (CCTC) loss to encourage spelling consistencies of a character-based non-autoregressive ASR which allows for faster inference.
The model trained by CCTC loss is aware of contexts since it learns to predict both center and surrounding letters in a multi-task manner.
In contrast to existing CTC-based approaches, CCTC loss does not require frame-level alignments, since the context ground truth is obtained from the model's estimated path.
Compared to the same model trained with regular CTC loss, our method consistently improved the ASR performance on both CS and monolingual corpora.

%% file: tex_files/intro.tex
\vspace{-2pt}
\section{Introduction}
\label{sec:intro}
\vspace{-4pt}
Code-Switching (CS) is a phenomenon in which multiple languages are used in a single conversation. CS can appear on word, phrase, or sentence level and can alternate throughout the conversation. If the language changes between sentences, it is called inter-sentential CS. Instead, if CS happens within a single utterance or sentence, it is called intra-sentential CS, which is the focus of this work. 

Connectionist Temporal Classification (CTC) loss \cite{graves2006connectionist} has gained popularity in the community due to its computational efficiency and ability to train character-based end-to-end models. Character-based models, requiring no dictionary, lead to greater usability across different languages in both mono- \cite{miiller2018multilingual, tong2018cross, wang2019end, dalmia2018domain, ito2017end} and multilingual \cite{li2019towards,sreeram2020exploration, yue2019end} setups.
However, when dealing with utterances with CS, 
an outputted word from a CTC model can contain a mixture of characters from different languages because the model lacks the context when predicting the output.
To avoid this problem, 
multilingual sentences are transliterated into one language \cite{emond2018transliteration} which can be unnatural.

The lack of context can be alleviated by postprocessing the prediction using a Language Model (LM) or providing context information from past states. However, LM rescoring increases the inference time. Similarly, Long Short-Term Memory (LSTM) or previous hidden states can be used to provide context at the expense of inference speed. One can also add language information to the model which can be done jointly with the main ASR task in a multi-task manner. This language identification subtask has been explored either on the frame level \cite{li2019towards, luo2018towards} or the subword level \cite{Zeng2019, shan2019investigating}. However, adding a language identifier to  CTC model requires alignments since those identifiers are attached to every timestep of the model's prediction.

For monolingual ASR, teaching the model to predict future context has led to improvements in the main prediction. 
In \cite{Jaitly2014AutoregressivePO}, the model was trained to predict multiple frames at once given a contextualized input. 
In \cite{zhang2015speech,zhang2016prediction}, the main prediction was based on both the current frame features and the predictions made on previous frames.
We hypothesize that providing the required contextual information to the model in this manner should also alleviate the inconsistency issue in CS utterances without sacrificing speed. 
Nevertheless, predicting future context in the CTC framework is not straightforward since there is no explicit frame-level alignment for computing the Cross-Entropy (CE) loss.
Besides, outputs of the CTC model are mostly blanks, providing little contextual information.

In this work, we propose Contextualized CTC (CCTC) loss, a modification to the original CTC-based approach that incorporates awareness of the nearby letters.
Chorowski et al. \cite{Chorowski2019} used tri-characters as prediction outputs to incorporate context information to CTC. This requires a change to the decoder in order to support context-dependent outputs. In our approach, we use single characters as output targets and have the model predicts the surrounding letters as well. This is done by adding secondary prediction heads to predict the surrounding characters in a multi-task manner.
The novelty of our approach lies on how we obtain the labels for the surrounding characters.
The target prediction for the surrounding characters is obtained by the prediction from the previous iteration which replaces the need for frame-based alignments from Hidden-Markov-Model-based or Attention-based models. To get the prediction target of the context we also ignore blanks and consecutive duplicate letters. This helps provide longer contextual information than predicting actual symbols of surrounding frames, which has shown to be more effective in \cite{zhang2015speech}.

Our experiments on a Thai-English (TH-EN) CS corpus show that the CCTC loss can help mitigate intra-word language inconsistency. We show that applying the CCTC loss has the same effect as implicitly learning a low-order character LM, yet gives complementary gains when combined with word-based n-gram rescoring. 

%% file: tex_files/background.tex
\section{Background}
\label{sec:background}
In this section, we give an overview of the CTC loss \cite{graves2006connectionist}.
CTC is an alignment-free objective function for sequence-to-sequence tasks such as ASR and handwriting recognition. 
Suppose we have an input sequence, $x = (x_1, x_2, ..., x_T)$, 
CTC loss maximizes the probability of predicting the ground truth transcription, $y^* = (y^*_1, y^*_2, ..., y^*_U): y^*_t \in \mathbf{A}$, where $\mathbf{A}$ is a set containing alphabets in the languages which in this work are Thai and English alphabets.
A blank token, $\epsilon$, is also included in the alphabets set, $\mathbf{A}' = \mathbf{A} \cup \{ \epsilon \}$, to handle noise, silence, and consecutive duplicate characters in transcriptions.
CTC model outputs a path, $\pi = (\pi_1, \pi_2, ..., \pi_T) : \pi_t \in \mathbf{A'}$, which has the same length as the input frames.
Lastly, $\pi$ is mapped to an inferred transcription, $y = (y_1, y_2, ..., y_K) : y_k \in \mathbf{A}$, using a mapping function $\mathcal{B}: \mathbf{A}'\to \mathbf{A}$.
By applying the function $\mathcal{B}$, adjacent duplicate alphabets are merged and blank tokens are removed.

The CTC loss, $\mathcal{L}_{CTC}$, is the negative log probability of all paths that can be mapped to the ground truth transcription. The CTC loss is calculated as follows: 

\begin{equation}
    \label{eqn:ctc_y}
    P(y | x) = \sum_{\pi \in \mathcal{B}^{-1}(y)} P(\pi | x) = 
    \sum_{\pi \in \mathcal{B}^{-1}(y)} \prod_t P(\pi_t|x)
\end{equation}

\begin{equation}
    \label{eqn:ctc}
    \mathcal{L}_{CTC} = - \log P(y | x)
\end{equation}

\begin{figure}[t]
  \centering
  \includegraphics[width=.95\linewidth]{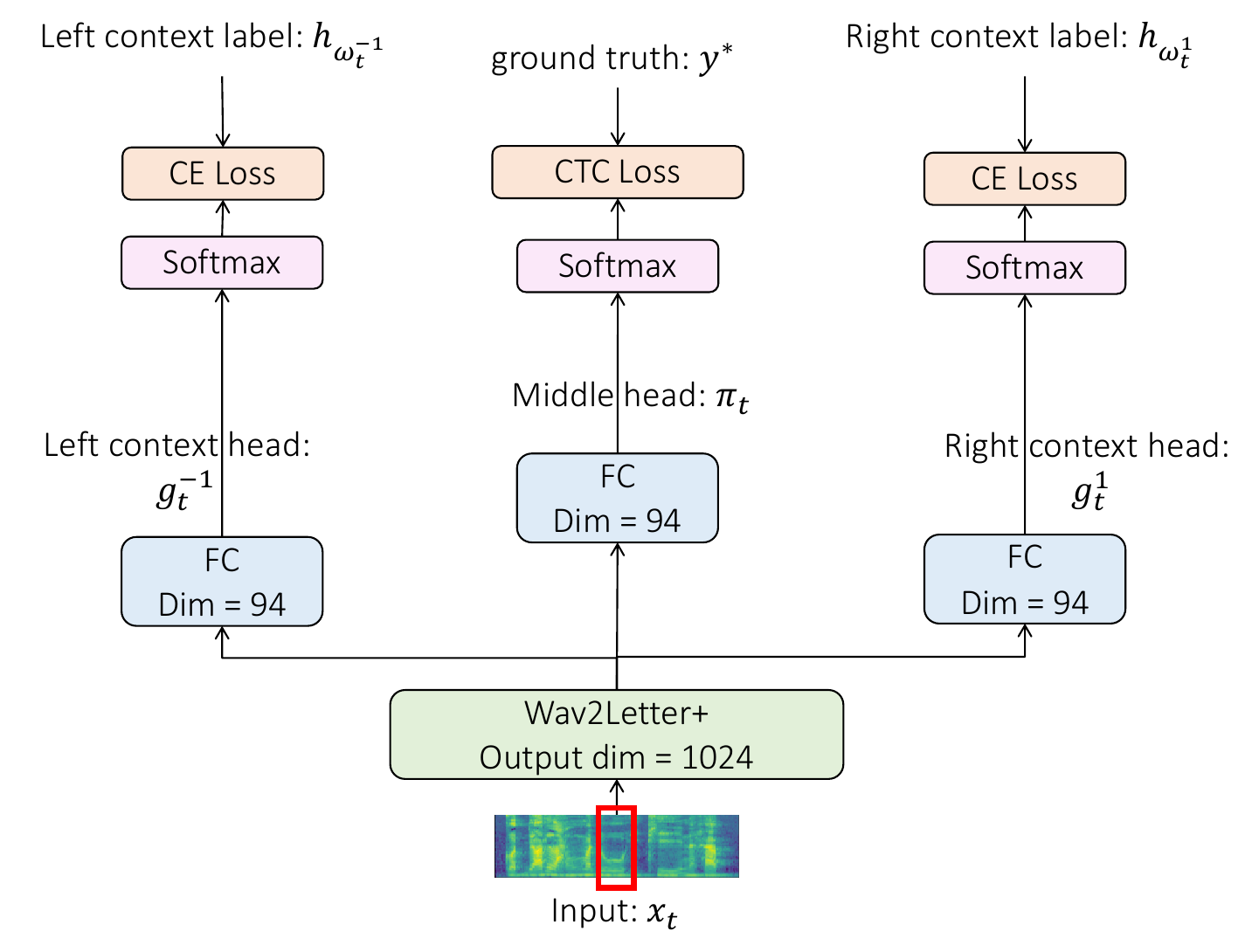}
  \caption{The CCTC architecture used in this work. Wav2Letter+ is modified by adding two extra prediction heads. 
  The model is now aware of contexts as it learns to predict previous, current, and next characters simultaneously.
  }
  \label{fig:model}
\end{figure}

\begin{figure}[t]
  \centering
    \includegraphics[width=\linewidth]{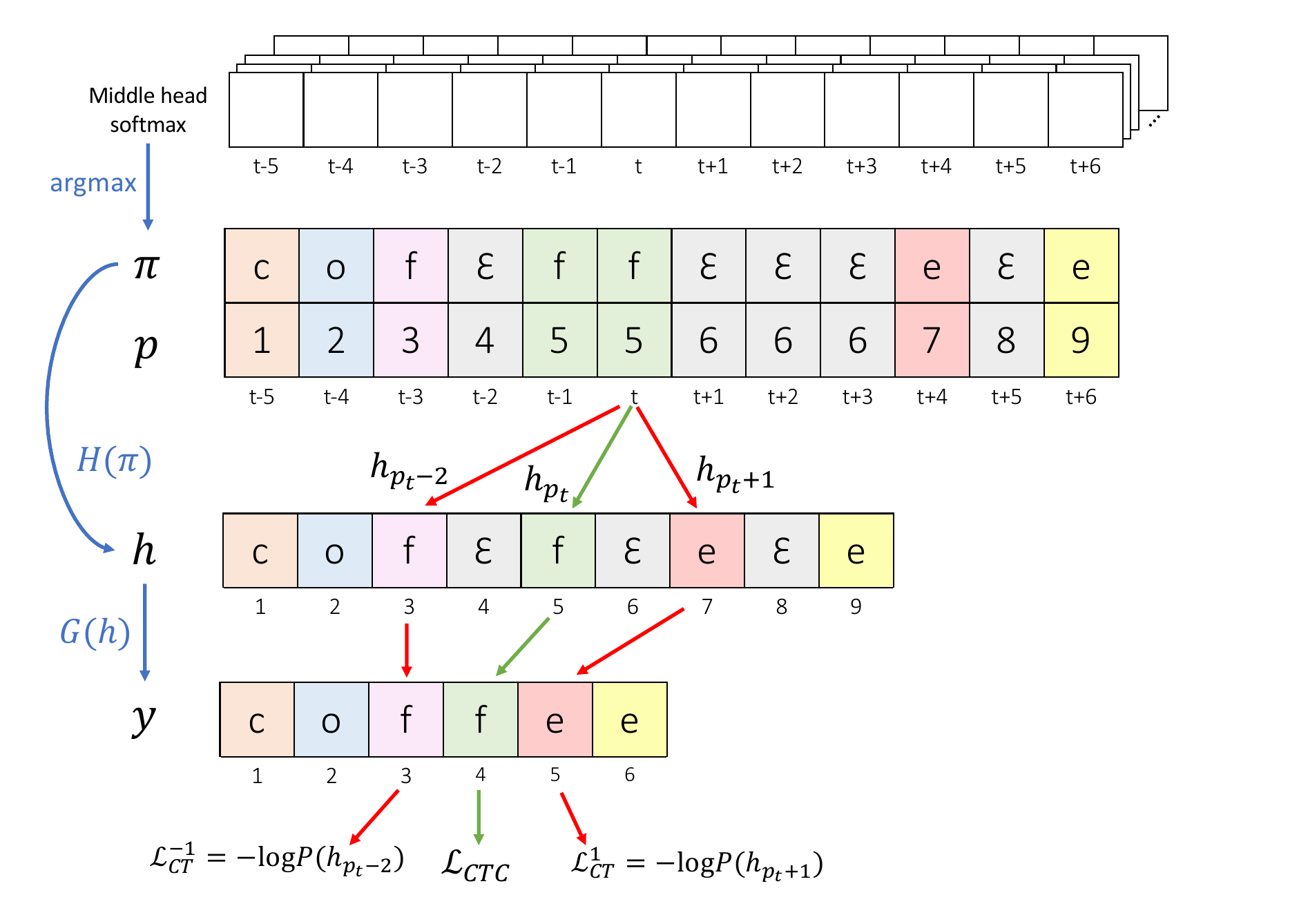}
  \caption{Label generation procedure for the CCTC context loss. Given a path from the output of the middle head's softmax, the labels for the left and right heads can be found by applying the functions $H$ and $G$. For frame index $t$ the left and right targets are f and e, respectively. 
  The correspondence between $\pi$ and $y$ are color coded. A letter, $y_i$, is derived from the token, $\pi_t$, with the same background color.
}
  \label{fig:loss}
\end{figure}

%% file: tex_files/method.tex
\section{Contextualized Connectionist Temporal Classification}
\label{sec:method:CCTC}
The motivation for the CCTC loss is to indirectly introduce context conditioning to the ASR models focusing on speed, which usually are non-autoregressive (NAR) and non-recurrent.
Since a NAR model is unaware of the surrounding predictions during inference, we introduce dependencies on the predicted contexts, $g_t$, to encourage the model to generate more consistent predictions.
For an input $x_t$, the model predicts both the output token, $\pi_t$, and its contexts, $g_t = (g_t^{-K}, ..., g_t^{-2}, g_t^{-1}, g_t^{1},  g_t^{2}, ..., g_t^{K})$, as shown in Fig.~\ref{fig:model}.
We define $K$ as the size of context and use the superscript, $k$, of a context, $g_t^k$, to indicate the position of the context relative to the token $\pi_t$.
We refer to the CCTC loss with the context size $K=1$ as a $1^{st}$-order CCTC from now on.
The context prediction heads minimize the CE loss between the predicted contexts, $g_t$, and their ground truths which are the characters on the path that will be predicted by the middle head. 
The middle head minimizes the regular CTC loss.
%
%
Hence, the model learns probability distributions of the center letter, $P(\pi_t|x_t)$, and surrounding contexts, $P(g_t|x_t)$, at the same time.

Since a typical CTC path, $\pi$, usually contains a lot of blank tokens which is not informative, we opt to train the context heads with dense character supervision from the prediction, $y = \mathcal{B}(\pi)$, instead.
Concretely, the predicted contexts, $g_t^{-k}$ and $g_t^{k}$, at a position $t$ are the k\textsuperscript{th}-nearest characters to the left and right of $\pi_t$ that are not a blank token or consecutive duplicate.
However, a naive search of labels for every context, $g_t$, is computationally expensive.
Thus, we propose the following efficient and generalizable algorithm that could scale to any order of the CCTC loss on top of the existing CTC pipeline.

To obtain the ground truths, we decompose the mapping function $\mathcal{B}$ into two steps, namely removing blank tokens and removing duplicates. We define $H$ as a mapping that merges all consecutive duplicates into one and $G$ as a mapping which removes blank tokens from the sequence.
Thus, the path mapping $\mathcal{B}$ can be written as $\mathcal{B}(\pi) = G(H(\pi))$.
Given a merged path, $ h=(h_1, h_2, ..., h_L) = H(\pi)$, we create a list, $p = (p_1, p_2, ..., p_T) : p_t \in [1,L], p_t \leq p_{t+1} $, containing indices of $h$. An index, $p_t$, indicates that a letter, $h_{p_t}$, is derived from a path token, $\pi_t$, after applying the merging function $H$.
We demonstrate this process with an example in Fig.~\ref{fig:loss}.
As a result, we can use indices, $p_t-1$ and $p_t+1$, to identify nearest letters, $h_{p_t-1}$ and $h_{p_t+1}$, that are not a consecutive duplicate of $\pi_t$.

To calculate the $1^{st}$-order context losses, $\mathcal{L}_{CT}^{-1}$ and  $\mathcal{L}_{CT}^{1}$, for training $1^{st}$-order context heads, we use (\ref{eqn:CCTC_1_omg}) to 
compute indices, $\omega_t^{-1}$ and $\omega_t^1$, which indicate the first non-blank letters to the left and the right of the middle letter $h_{p_{t}}$.
\begin{equation}
    \label{eqn:CCTC_1_omg}
    \omega_t^k =
    \begin{cases}
     p_{t+d}, & \text{if $h_{p_{t+d}} \neq \epsilon$}.\\
     p_{t+2d}, & \text{otherwise}.
\end{cases}
\end{equation}
where the superscript $k \in \{-1, 1\}$ denotes the order of the context loss. The variable $d$ is $-1$ when $k < 0$ and $1$, otherwise.
Then, we apply the CE criterion on the obtained labels, $h_{\omega_t^{-1}}$ and $h_{\omega_t^{1}}$ , as shown in (\ref{eqn:cctc_loss}).
\begin{equation}
\label{eqn:cctc_loss}
 \mathcal{L}_{CT}^{k} = - \sum_t{ \log P( h_{\omega^{k}_t} ) }
\end{equation}


For a wider context: $1<|k|<=K$, we can compute the $k^{th}$-order context losses, $\mathcal{L}_{CT}^{-k}$ and $\mathcal{L}_{CT}^{k}$, recursively based on the previously found index $\omega_t^{k-d}$ using (\ref{eqn:CCTC_k_omg}). 
\begin{equation}
  \label{eqn:CCTC_k_omg}
  \omega^k_t = 
  \begin{cases}
    \omega^{k-d}_t+d, & \text{if $h_{\omega^{k-d}_t+d} \neq \epsilon$}.\\
    \omega^{k-d}_t+2d, & \text{otherwise}.
  \end{cases} 
\end{equation}

Finally, the CCTC loss is the combination of the CTC loss and the context losses up to the $K^{th}$ order as shown in (\ref{eqn:loss}). The weights of the left and right contexts can be set differently with $\alpha$ and $\beta$.
\begin{equation}
    \mathcal{L}_{CCTC}^{K} = \mathcal{L}_{CTC} + \sum_{k=1}^K \alpha^{-k} \mathcal{L}^{-k}_{CT} + \beta^{k} \mathcal{L}^{k}_{CT} 
    \label{eqn:loss}
\end{equation}

%% file: tex_files/dataset.tex
\section{Dataset}
\label{sec:dataset}
For our experiments, we used a 200-hour Thai speech corpus, crawled from public YouTube podcast channels. The utterances were then manually transcribed. 
The recordings were preprocessed to 16kHz and 16-bit depth.
CS with English was found in 4.4\% of the training set, 4.3\% of the development set, and 5.7\% of the test set.
More details are shown in Table~\ref{tab:dataset}. 
The YouTube channels in the test set are different from the training and development sets. Therefore, speakers in the test set are not in the training data.

For performance comparison and analysis, we separated the development set and test set into the monolingual part containing only Thai utterances and the CS part. We refer to these subsets as TH and TH-CS from now on. Note that training and hyperparameter tuning were still done on the entire set, making no such distinction.

\begin{table}[th]
\centering
\begin{tabular}{l| c c c c c }
 \hline 
    & Train & Development & Test\\ 
 \hline
duration		           & 150 Hr    &24 Hr    &26 Hr \\
\#total utterances         & 190K      &30K      &35K   \\
\#TH-CS utterances     & 8.4K      &1.3K     &2K   \\
\#TH letters		       & 7M        &1M       &1M    \\
\#EN letters               & 84K       &30K      &19K    \\
\#TH words          & 1.9M      &293K     &333K  \\
\#EN words          & 14K       &2K       &3K  \\
\#TH vocabulary      & 36K       &12K      &13K   \\
\#EN vocabulary      & 3K       &1K      &1K   \\
 \hline
\end{tabular}
\caption{Statistics of the dataset used in this work}

\label{tab:dataset}
\end{table}
\vspace{-10pt}

%% file: tex_files/exp.tex
\section{Experiments and Results}
We conducted a series of experiments to measure the performance of models trained with our proposed CCTC loss.
The experiments are designed to compare the CCTC loss with the standard CTC loss.
The details of our implementation are provided in Sec.~\ref{sec:result:implemetation}.
We present the results on our TH-EN dataset in Sec.~\ref{sec:result:en_th}, and on the LibriSpeech dataset in Sec.~\ref{sec:result:libri}.
We also show the effect of different CCTC loss weights in Sec.~\ref{sec:result:ct_weight} and different sizes of beam width in Sec.~\ref{sec:result:beam_widths}.

\subsection{Experimentation details}
\label{sec:result:implemetation}

\input{tex_files/exp/detail}


\input{tex_files/exp/enth}

\subsection{Effect of CCTC loss on Monolingual English}
\label{sec:result:libri}

\input{tex_files/exp/libri}

\subsection{Effect of CCTC context loss weight}
\label{sec:result:ct_weight}
\input{tex_files/exp/ct_weight}


\subsection{Effect of beam width}
\label{sec:result:beam_widths}
\input{tex_files/exp/beam_width2}

%% file: tex_files/exp/detail.tex
We adopted a NAR and fully-convolutional model Wav2Letter+ \cite{kuchaiev2018mixed}, a modified version of Wav2Letter \cite{collobert2016wav2letter, liptchinsky2017based}, as our base model.  It comprises of 17 1D-convolutional layers and two fully-connected layers at the end.
We added context prediction heads right after the last layer of the base model as shown in Fig.~\ref{fig:model}. 
For simplicity, we only considered $1^{st}$-order CTCC. We also set $\alpha = \beta$ and tuned them using the development set.

Since the labels for the context heads are derived from the predicted path of the middle head, it is important that the context losses are applied only when these predictions are reliable. Therefore, in all experiments, we started by training the models with only the CTC loss for 300 epochs. Afterwards, the context losses were included, and the training resumed for an additional of 100 epochs. This additional training was also performed on the CTC baseline models. For each mini-batch, the context labels were generated on-the-fly with the current model's output path for efficiency.

The default settings of Wav2Letter+\footnote{ \url{https://github.com/NVIDIA/OpenSeq2Seq} } were used with some exceptions. The Adam optimizer \cite{DBLP:journals/corr/KingmaB14} was used instead of the original SGD optimizer.
The Layer-wise Adaptive Rate Control \cite{you2017large} and weight decay were not used as we found them hurting the performance. 
We replaced the polynomial decay with an exponential decay with a rate of 0.98.

LM rescoring was also applied to investigate more realistic setups. We curated two corpora with 27M words/145M letters from Thai Wikipedia and 69M words/330M letters from Pantip (Thai Q\&A forum). 
For each corpus, we did word tokenization using DeepCut \cite{Kittinaradorn2019} and trained word-based n-gram models using KenLM \cite{Heafield-kenlm}.
The final LM is obtained by n-gram interpolation. A beam width of 64 was used for LM rescoring unless stated otherwise.

At training time, context heads and the middle head were jointly optimized in a multi-task manner. 
At inference time, context heads were removed to preserve memory usage and computational cost since transcriptions were inferred by the middle head only.

%% file: tex_files/exp/enth.tex
\begin{table}[th]
\centering
\begin{tabular}{l|l||c c c}
  \hline
  \multirow{2}{*}{Data} 
    & \multirow{2}{*}{Model} 
      & \multicolumn{3}{c}{WER (\%)}  \\
          \cline{3-5}
  & &argmax & beam & 3-gram  \\  \hline
\multicolumn{5}{c}{Development set}\\
\hline
\multirow{2}{*}{TH} & CTC  & 15.01             &  14.89                & 13.27                \\  
                    & CCTC & \textbf{14.67}*    &  \textbf{14.58}*               & \textbf{13.17}*     \\ 

\hline
\multirow{2}{*}{TH-CS} &CTC	  &28.02           &  27.76                & 24.09             \\
                       &CCTC  &\textbf{27.57}*  & \textbf{27.43}                 & \textbf{23.78}   \\ 
\hline
\multicolumn{5}{c}{Test set}\\
\hline
\multirow{2}{*}{TH} & CTC  & 15.66             &  15.52                & 13.47              \\
                    & CCTC & \textbf{15.30}*    & \textbf{15.22}*                &\textbf{13.42}     \\ 
\hline
\multirow{2}{*}{TH-CS} &CTC	 &27.73            &   27.74               & 25.04               \\  
                       &CCTC &\textbf{27.33}*   &  \textbf{27.27}*               &\textbf{24.56}*     \\ 
\end{tabular}
\caption{The evaluation on development and test sets. The * symbol indicates a significant difference at $p < 0.05$ to the baseline CTC using MAPSSWE two-tailed tests.}
\label{tab:CCTC_cer}
\end{table}
 
\subsection{Effect of CCTC loss on EN-TH dataset}
\label{sec:result:en_th}
In the CS dataset setup, we investigated the performance of word recognition on the model trained by CCTC loss compared to standard CTC loss using a standard metric: Word Error Rate (WER).

Table~\ref{tab:CCTC_cer} shows that CCTC loss outperforms CTC loss in both with and without LM setups.
In the monolingual TH test set, we found that the more context we provided to the decoder the lesser performance gains we observed. 
The relative improvements of CCTC over CTC decreased from 2.3\% to 1.9\% and 0.4\% when we changed from an argmax decoder to beam search decoder without LM and with LM, respectively.
In contrast, we found the opposite trend in the TH-CS test set. 
The relative improvements steadily increased from 1.4\% to 1.7\% and 2.0\%. These complementary gains were an effect of having both short (CCTC) and long (LM) context information.

\begin{figure}
  \centering
  \centerline{\includegraphics[width=\linewidth]{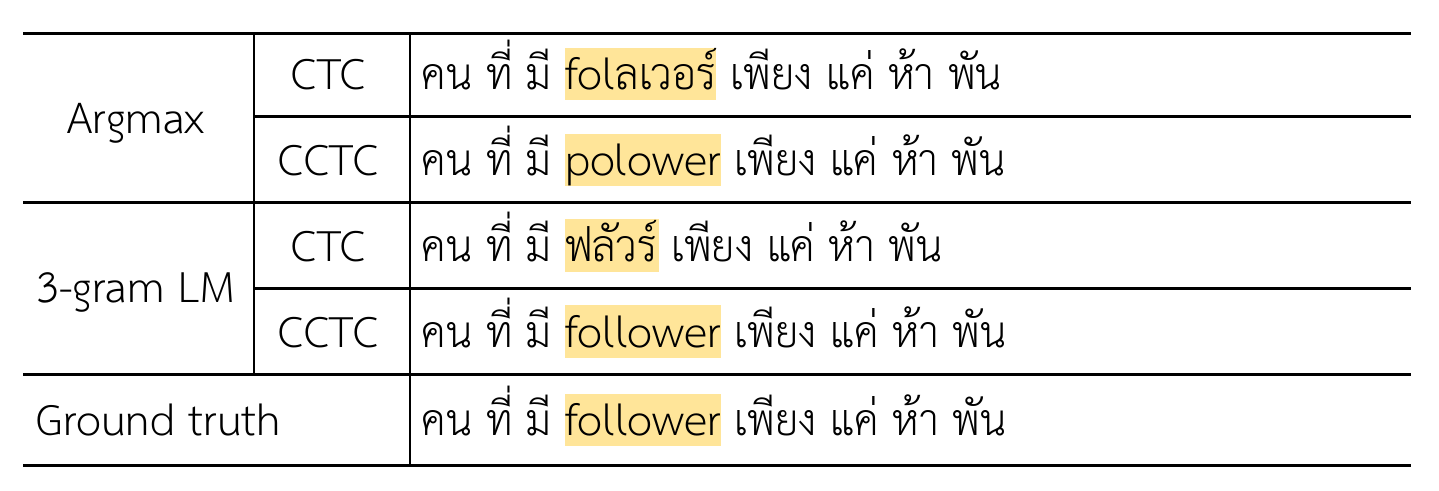}}
  \caption{Comparison between models for selected examples. Aligned differences are highlight in different colors.}
  \label{fig:sample}
\end{figure}

Further qualitative analysis shows that CTCC mostly fixes the inconsistencies in the spelling. Fig.~\ref{fig:sample}. depicts the word ``follower'' spelled with a mixture of Thai and English alphabets in the CTC model, while CCTC model outputs English alphabets consistently. The phoneme sequence /ol/ only appears in loanwords in Thai, making the model heavily prefers to output ``ol.''

Note that spelling inconsistencies still exist after applying LM rescoring for both CTC and CCTC models. However, CCTC reduces the number of words spelled with both Thai and English characters by half, proving the usefulness of CCTC for CS ASR. To completely remove the inconsistent words, one can increase the LM weight. However, this can cause issues for out-of-vocabulary words.



We also attempted to train a Listen-Attend-Spell (LAS) model \cite{chan2016listen,liu2020sequence} as a representative autoregressive model. 
The results on TH and TH-CS dev sets were 16\% and 30\% WER, much worse than the CTC baseline. This is consistent with other works that were trained on data of similar size \cite{liu2020sequence, watanabe2017hybrid}\footnote{We trained the model on LibriSpeech 100 hours for the comparison.}, 
Thus, we decided not to include the results in the comparison as LAS models seem to require larger amounts of training data to perform well.

%% file: tex_files/exp/libri.tex
In this experiment, we investigated the performance of CCTC loss on a 100 hours subset of LibriSpeech \cite{panayotov2015librispeech}. The goal of this experiment is to determine whether CTCC is beneficial in other languages. 

Table~\ref{tab:libri_result} summarizes WER on the test-clean utterances. We provided results from the Wav2Letter++ \cite{pratap2019wav2letter++} model taken from the Wav2Letter tutorial\footnote{\url{https://github.com/facebookresearch/wav2letter/tree/recipes-conv-glu-paper/tutorials/1-librispeech\_clean}} which was trained on the same subset as a strong baseline.
The CCTC model improves the WER over the CTC model by 3.1\% relative when used with an argmax decoder. 
Although the improvement is small when LM rescoring is applied, it still suggests that applying CCTC loss consistently yields performance gain over CTC loss without losing inference speed.

\begin{table}
\centering
\begin{tabular}{l| l | c}
 \hline
Model & Decoder & WER \\ 
\hline
Wav2Letter+ w/ CTC                                          & argmax                             &22.00  \\
Wav2Letter+ w/ CCTC                                         & argmax                             &\textbf{21.32}  \\
\hline
Wav2Letter++ w/ CTC \cite{pratap2019wav2letter++}   & beam w/ 3-gram LM                   &18.97  \\
Wav2Letter+ w/ CTC                                          & beam w/ 3-gram LM                   & 15.72\\
Wav2Letter+ w/ CCTC                                         & beam w/ 3-gram LM                   & \textbf{15.67}\\
\end{tabular}
\caption{The evaluation on LibriSpeech test-clean set.}
\label{tab:libri_result}
\end{table}

%% file: tex_files/exp/ct_weight.tex
\begin{figure}
  \centering
  \includegraphics[width=\linewidth]{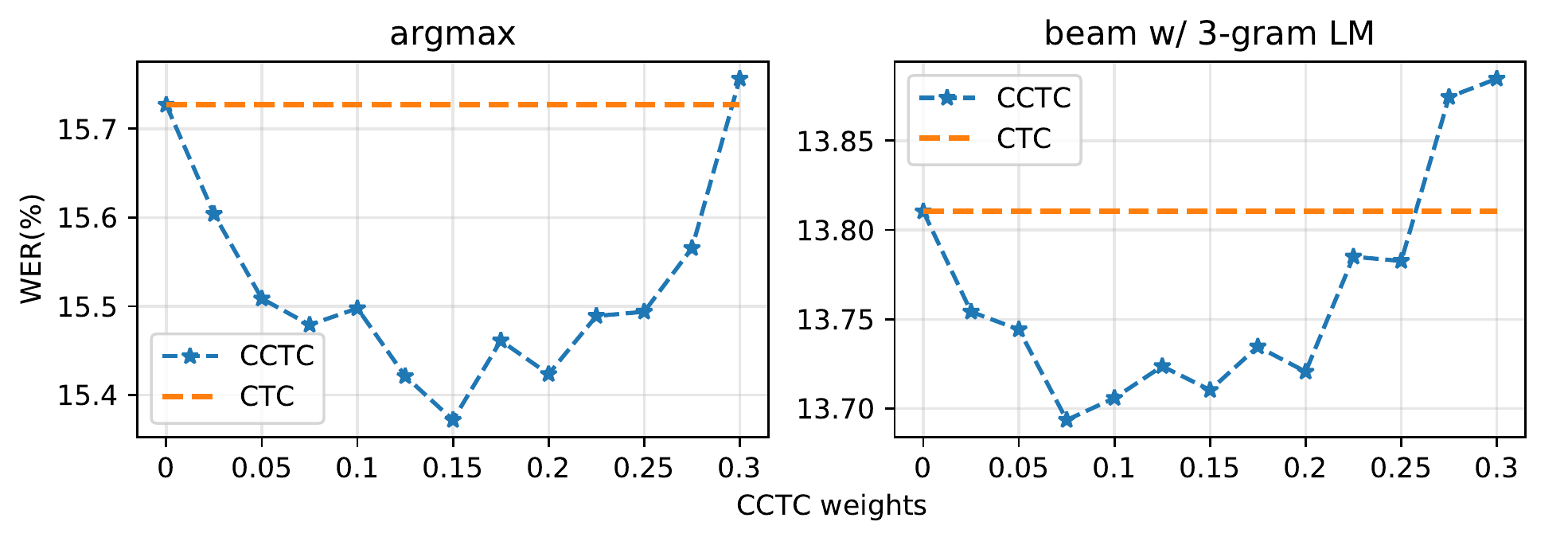}
  \caption{WER on the dev set as the context loss weight changes}
  \label{fig:cctc_weights}
\end{figure}

As the weights of the context losses, $\alpha$ and $\beta$, control the trade-off between the generated contexts and the middle prediction during training, we studied how the weights can affect the performance.
Fig.~\ref{fig:cctc_weights} summarizes the results on the entire (TH and TH-CS) dev set.

The optimal context loss weight is larger when a greedy decoder is used compared to a beam search decoder. This is due to the fact that when the external LM is not available, the model needs to rely more on the context heads to make consistent predictions. 
In general, we found any value between 0.1-0.2 yields improvement over regular CTC on both LibriSpeech and our Thai corpus.

\vspace{-5pt}

%% file: tex_files/exp/beam_width2.tex
\begin{figure}
  \centering
  \includegraphics[width=\linewidth]{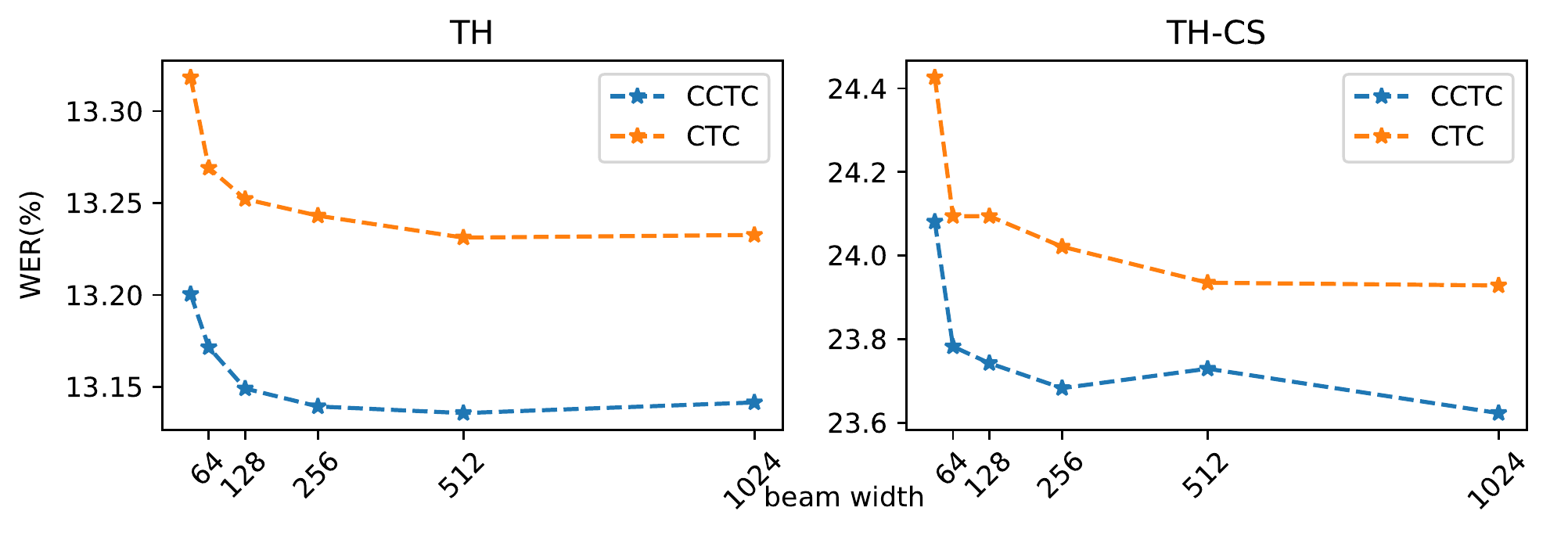}
  \caption{Performance comparison between CTC and CTCC using different size of beam for LM rescoring}
  \label{fig:beam_width_graph}
\end{figure}

We studied the correlation between beam width and the WER of both CTC and CCTC models on the dev set.
We did the experiment by varying the beam size from 32 up to 1024.
Fig.~\ref{fig:beam_width_graph} shows that CCTC consistently improves the WER of both TH and TH-CS dev sets.
The gain from CTCC loss and LM rescoring seems to be complementary regardless of the size of beam.


%% file: tex_files/conclusion.tex
\section{Conclusions}
We introduced CCTC loss for incorporating context information into a CTC-based NAR model without increasing inference time.
We showed that the CCTC loss improved results in utterances with  CS by encouraging context consistency in the predicted path.
We believe that the technique of adding additional context dependencies in the CCTC loss can be helpful for CS ASR regardless of the language.
In the future, we plan to investigate the impact of different left and right context weights as it might be more natural for the model to depend more on the preceding characters.
We also plan to measure the effect of incorporating larger context sizes into the CCTC loss.